\documentclass[a4paper,UKenglish,cleveref, autoref, thm-restate]{lipics-v2021}

\pdfoutput=1 
\hideLIPIcs  

\title{A formalization of the Gelfond-Schneider theorem} 

\author{Michail Karatarakis}{Institute for Computing and Information Sciences, Radboud University, Nijmegen, The Netherlands\and \url{https://www.cs.ru.nl/staff/Michail.Karatarakis} }{michail.karatarakis@ru.nl}{https://orcid.org/0000-0002-4118-6206}{}

\author{Freek Wiedijk}{Institute for Computing and Information Sciences, Radboud University, Nijmegen, The Netherlands\and \url{https://www.cs.ru.nl/staff/Freek.Wiedijk/} }{freek@cs.ru.nl}
{[orcid]}
{}

\authorrunning{Michail Karatarakis and Freek Wiedijk}

\Copyright{Michail Karatarakis and Freek Wiedijk} 



\category{} 

\relatedversion{} 

\acknowledgements{This project began during the 2024 Prospects of Formal Mathematics program held at the Hausdorff Institute for Mathematics (HIM). We are grateful to the organizers for creating a collaborative environment and to the HIM for their hospitality. Additionally, we extend our thanks to Mario Carneiro, Raphael Douglas Giles, and the Mathlib reviewers.}
\nolinenumbers 

\EventEditors{John Q. Open and Joan R. Access}
\EventNoEds{2}
\EventLongTitle{42nd Conference on Very Important Topics (CVIT 2016)}
\EventShortTitle{CVIT 2016}
\EventAcronym{CVIT}
\EventYear{2016}
\EventDate{December 24--27, 2016}
\EventLocation{Little Whinging, United Kingdom}
\EventLogo{}
\SeriesVolume{42}
\ArticleNo{23}
\usepackage{listings}
\usepackage{upgreek}
\usepackage{tikz}
\usepackage[noend]{algpseudocode}
\usepackage{float}
\usetikzlibrary{automata, positioning}
\usepackage{amsmath}
\usepackage{adjustbox}
\usepackage{hyperref} 
\usepackage{xcolor}
\usepackage{csquotes}
\usepackage{todonotes}
\newtheorem{thm}{Theorem}[section]

\usepackage{xcolor}
\definecolor{keywordcolor}{rgb}{0.7, 0.1, 0.1}   
\definecolor{tacticcolor}{rgb}{0.0, 0.1, 0.6}    
\definecolor{commentcolor}{rgb}{0.4, 0.4, 0.4}   
\definecolor{symbolcolor}{rgb}{0.0, 0.1, 0.6}    
\definecolor{sortcolor}{rgb}{0.1, 0.5, 0.1}      
\definecolor{attributecolor}{rgb}{0.7, 0.1, 0.1} 

\def\aroof{\overline{\hspace{-1pt}\smash[t]{|}\hspace{1pt}\alpha\hspace{1pt}\smash[t]{|}\hspace{-1pt}}}

\definecolor{academicpurple}{rgb}{0.4,0.0,0.35}
\newcommand{\secref}[1]{\hyperref[#1]{\textcolor{academicpurple}{§\ref*{#1}}}}

\newcommand{\citeref}[1]{\textcolor{academicpurple}{\cite{#1}}}

\newcommand{\thmref}[1]{\textcolor{academicpurple}{\hyperref[#1]{Theorem~\ref*{#1}}}}

\newcommand{\eqrefcol}[1]{\textcolor{academicpurple}{\hyperref[#1]{Equation~(\ref*{#1})}}}

\definecolor{academicpurple}{rgb}{0.4,0.0,0.35}

\usepackage{xcolor}
\definecolor{keywordcolor}{rgb}{0.7, 0.1, 0.1}   
\definecolor{tacticcolor}{rgb}{0.0, 0.1, 0.6}    
\definecolor{commentcolor}{rgb}{0.4, 0.4, 0.4}   
\definecolor{symbolcolor}{rgb}{0.0, 0.1, 0.6}    
\definecolor{sortcolor}{rgb}{0.1, 0.5, 0.1}      
\definecolor{attributecolor}{rgb}{0.7, 0.1, 0.1} 

\newcommand{\extlink}{~\ensuremath{{}^\text{\faExternalLink*}}}
\def\aroof{\overline{\hspace{-1pt}\smash[t]{|}\hspace{1pt}\alpha\hspace{1pt}\smash[t]{|}\hspace{-1pt}}}

\makeatletter
\def\do@url@hyp{\do\-}
\makeatother

\makeatletter
\def\do@url@hyp{\do\-}
\makeatother

\newcommand{\roof}[1]{\overline{\hspace{-1pt}\smash[t]{|}\hspace{1pt}#1\hspace{1pt}\smash[t]{|}\hspace{-1pt}}}

\begin{document}

\maketitle


\begin{abstract}
We formalize Hilbert's Seventh Problem and its solution, the Gelfond-Schneider theorem, in the Lean 4 proof assistant. The theorem states that if $\alpha$ and $\beta$ are algebraic numbers with $\alpha \neq 0,1$ and $\beta$ irrational, then $\alpha^\beta$ is transcendental. Originally proven independently by Gelfond and Schneider in 1934, this result is a cornerstone of transcendental number theory, bridging algebraic number theory and complex analysis.

\end{abstract}

\section{Introduction}
The classification of real numbers is a fundamental theme in number theory. Beyond the distinction between rational and irrational numbers lies the deeper partition into \emph{algebraic} and \emph{transcendental} numbers. An algebraic number is defined as a root of a non-zero polynomial with integer coefficients; examples include rational numbers like $1/2$ or surds like $\sqrt{2}.$ Conversely, a transcendental number satisfies no such polynomial relation. While set-theoretic arguments demonstrate that almost all real numbers are transcendental, the scarcity of provable examples in the literature highlights the difficulty of establishing transcendence for specific constants. 

The first constructive proof of the existence of transcendental numbers was provided by Joseph Liouville in 1844~\citeref{liouville1844}. By analyzing how well algebraic numbers can be approximated by rationals, Liouville constructed a specific class of transcendental numbers. However, his methods were insufficient to determine the nature of classical constants such as $e$ and $\pi,$ or to address exponential relationships. A major breakthrough occurred in the late 19th century with the introduction of \emph{auxiliary functions} -- functions explicitly constructed to vanish to high order at points of interest. Using this technique, Hermite established the transcendence of $e$ in 1873~\citeref{hermite1873}. Lindemann refined the method in 1882 to prove the Lindemann--Weierstrass theorem, which states that for any distinct algebraic numbers $\alpha_1, \dots, \alpha_n$, the exponentials $e^{\alpha_1}, \dots, e^{\alpha_n}$ are linearly independent over the rationals. A celebrated corollary~\citeref{lindemann1882} of this theorem is the transcendence of $\pi,$ resolving the ancient problem of squaring the circle. Despite these successes, the Lindemann--Weierstrass theorem did not address numbers of the form $\alpha^\beta$ where both the base and exponent are algebraic, such as $2^{\sqrt{2}}.$ This gap was settled by Hilbert's Seventh Problem. 

The study of transcendental numbers received significant impetus from David Hilbert when, at the 1900 International Congress of Mathematicians in Paris, he presented a famous list of twenty-three outstanding unsolved mathematical problems. Specifically, the first part of Hilbert's Seventh Problem was to decide whether numbers of the form $a^b$ are algebraic or transcendental, given that $a$ and $b$ are algebraic numbers (excluding the trivial cases where $a=0,$ $a=1,$ or $b$ is rational, which yield algebraic results). Hilbert explicitly asked for proofs regarding the transcendence of specific examples, notably $2^{\sqrt{2}}.$ The difficulty of this problem was underscored by Hilbert himself, who believed its solution would likely not be available before those of problems like the Riemann Hypothesis (Hilbert's Eighth Problem) and Fermat's Last Theorem. Nevertheless, the problem was completely settled in 1934 by Alexander Gelfond and, independently, by Theodor Schneider. This pivotal result, now known as the Gelfond–Schneider Theorem, states that if $\alpha$ and $\beta$ are algebraic numbers with $\alpha \neq 0,1$ and $\beta$ irrational, then $\alpha^\beta$ is transcendental. The elegant, yet difficult, proof of the Gelfond–Schneider Theorem serves as a crucial milestone in transcendental number theory, marking the beginning of the application of non-elementary analytical methods to solve fundamental algebraic problems. This theorem provides a wide class of explicitly transcendental numbers and demonstrates the power of the methods developed in transcendental number theory. Additionally, it establishes the transcendence of values derived from complex exponentiation, most notably Gelfond's constant $e^{\pi} = (-1)^{-i}$ and the principal value $i^i = e^{-\pi/2}.$
In the 1960s, Alan Baker advanced the field further by providing bounds for linear forms in logarithms, a problem initially posed by Gelfond. Baker showed that if $\alpha_1, \dots, \alpha_n$ are nonzero algebraic numbers, and $\beta_1, \dots, \beta_n$ are algebraic numbers such that $1, \beta_1, \dots, \beta_n$ are linearly independent over $\mathbb{Q}$, then the number
$$\alpha_1^{\beta_1} \alpha_2^{\beta_2} \cdots \alpha_n^{\beta_n}$$is transcendental. This result not only solved long-standing problems in transcendental number theory but also had important applications in Diophantine equations, earning Baker the Fields Medal~\citeref{baker1966,baker1967}.

Transcendental number theory has thus evolved from a collection of isolated facts into an important branch of number theory with its own problems and methods. Beyond proving the transcendence of specific numbers, the field also studies quantitative problems, such as results that provide estimates for approximating algebraic numbers by rational numbers. These strands illustrate both the depth and breadth of the field: it contains basic problems with simple statements that have eluded solution for centuries -- such as whether $e+\pi$ is irrational -- while simultaneously drawing upon and contributing to profound techniques in algebra and analysis~\citeref{shafarevich}.

\paragraph*{Related Work.}
The formalization of transcendental number theory is still in its early stages. Existing mechanized results focus primarily on classical proofs of the transcendence of specific constants. Bingham~\citeref{bingham_formalizing_2011} formalized Hermite's proof that $e$ is transcendental in HOL~Light, providing the first machine-checked transcendence result of this kind. Bernard, Bertot, Rideau, and Strub~\citeref{bernard_formal_2015} later formalized transcendence proofs for both $e$ and $\pi$ in Rocq, combining tools from the Coquelicot~\citeref{coquelicot} and Mathematical Components libraries. In the domain of constructive mathematics, Blot formalized basic properties of algebraic numbers and a proof of Liouville's theorem within the C-CoRN library, marking an early step in the mechanization of transcendental number theory using Rocq \citeref{cornliouville}.
 
In the Isabelle/HOL ecosystem, the Archive of Formal Proofs (AFP) contains significant contributions to transcendental number theory. Fleuriot and Paulson developed early libraries for transcendental functions. Other efforts focused on elementary results, such as a formalization of the irrationality of $e^r$ based on Aigner and Ziegler's exposition in {\it Proofs from THE BOOK}~\citeref{fromthebook}, and a development of Liouville numbers that establishes the existence of transcendental numbers via rational approximation~\citeref{liouvilleisabelle}. More recent AFP entries such as Eberl's~\citeref{eberl,eberlpi} formalization of the transcendence of $e$ and $\pi,$ and the work of Koutsoukou-Argyraki, Li, and Paulson~\citeref{koutsoukou-argyraki_irrationality_2022} on transcendence and irrationality criteria for infinite series extended the analytic infrastructure further. The Hermite-Lindemann-Weierstrass theorem has been formalized~\citeref{eberllindemann}; following the approach taken by Bernard in Rocq, this work utilizes Baker's version of the theorem to establish the algebraic independence of exponentials, thereby proving the transcendence of constants such as $e$ and $\pi.$

Parallel efforts in Mizar by Watase~\citeref{watase_formal_2024, watase_formal_2024-1} formalized a variant of Hurwitz's proof of the transcendence of $e.$ 
Despite this importance, the formalization of deep results in transcendental number theory remains largely unexplored, with existing work concentrated on isolated classical theorems rather than the broader theoretical framework.

In Lean, the transcendence of Liouville numbers has been formalized by the Lean community, including proofs that Liouville numbers are irrational and technical lemmas on integer polynomials evaluated at rational points, which ultimately establish their transcendence. Similarly, the analytic part of the Lindemann–Weierstrass theorem has also been formalized in Lean. Despite this progress, no proof assistant currently includes a formalization of the Gelfond--Schneider theorem. 

In this paper we formalize in Lean~4 the Gelfond--Schneider theorem, establishing the transcendence of $\alpha^\beta$ for any algebraic $\alpha \neq 0,1$ and any irrational algebraic $\beta.$ Our development includes the construction of the analytic and algebraic machinery required for the proof, such as auxiliary function arguments, growth estimates, and the necessary ingredients from algebraic number fields and complex analysis. To the authors' knowledge, this is the first formalization of the Gelfond--Schneider theorem in any theorem prover, and one of the first mechanized proofs of a deep result in transcendental number theory.

\paragraph*{Implementation details}
In fact, the proof is constructive and the original project idea was to do it constructively. However, \texttt{mathlib} is a classical library where the law of excluded middle and the axiom of choice are used extensively.
Finally, the standard library axiomatically introduces the axiom of choice (Chapter 12,~\citeref{tpil}) that 
produces data from a proposition asserting its existence, but this is not computationally
interpretable, and definitions that use this have to be marked as \texttt{noncomputable}.
An important aspect of this project is to contribute to \texttt{mathlib}, and parts of the source code have already been integrated in it. The code corresponding to this paper is publicly available on a \texttt{mathlib} branch{
\href{https://github.com/mkaratarakis/mathlib4/tree/9710b82b966813f7664819bc5d25b82d209a05e7/Mathlib/NumberTheory/Transcendental/GelfondSchneider}{\extlink}
}.


\paragraph*{Structure of the paper}
In Section 2, we describe the prerequisites we use from algebra and number theory, specifically the house of an algebraic number and Siegel's Lemma. Section 3 presents the formalization of Hilbert's seventh problem, detailing the construction of the auxiliary function and the analytic upper bounds required for the proof. Section 5 discusses the formalization process and the application to the Gelfond-Schneider constant, and Section 6 concludes the paper.

\section{Algebraic number theory}
A central component in Gelfond's proof of the transcendence of $\alpha^\beta$ is the construction of an auxiliary function that vanishes at specific points. The existence of the coefficients of this function, which must be algebraic integers of bounded size, is guaranteed by \emph{Siegel's Lemma}. While classical versions of Siegel's Lemma apply to rational integers, Gelfond's proof requires a generalization to the ring of integers $\mathcal{O}_K$ of a number field $K.$ This chapter details the formalization of the \emph{house} of an algebraic number and the subsequent proof of Siegel's Lemma for number fields (corresponding to Lemma 8.2 in the standard text).

\subsection{The House of an Algebraic Number}
For a rational integer $n \in \mathbb{Z}$, the absolute value $|n|$ is a sufficient measure of size. However, for an algebraic integer $\alpha \in \mathcal{O}_K$, a small absolute value in one embedding does not preclude $\alpha$ from being large in another. The corresponding measure of size is the maximum modulus of all conjugates of $\alpha$, known as the \emph{house} of $\alpha.$

\begin{definition}
Let $K$ be an algebraic number field of degree $h$, and let $\beta_1, \ldots, \beta_h$ be an integer basis, so that every integer in $K$ has the unique representation $a_1\beta_1 + \cdots + a_h\beta_h$, where $a_1, \ldots, a_h$ are rational integers. We denote by $\aroof$ the maximum modulus of the conjugates $a^{(i)}\mbox{ }(1 \le i \le h)$ of $a$, that is, $$\aroof = \max_{1 \leq i \leq h} |a^{(i)}| {
\href{https://github.com/mkaratarakis/mathlib4/blob/9710b82b966813f7664819bc5d25b82d209a05e7/Mathlib/NumberTheory/NumberField/House.lean#L39}{\extlink}
}.$$
\end{definition}

In formalizing the house of an algebraic number we initially considered a definition based on the maximum modulus of the roots of the minimal polynomial (using \texttt{Finset.max'}). 

\begin{lstlisting}
abbrev house (α : K) : ℝ := 
   (Complex.abs '' rootSet (minpoly ℚ α) ℂ).toFinset.max' (hmax α)
\end{lstlisting}

However, this approach treats conjugates as an unordered collection, obscuring their linear relationships and making arithmetic bounds like $\overline{|\alpha + \beta|} \le \overline{|\alpha|} + \overline{|\beta|}$ 
tedious to derive. To resolve this, we followed a structural suggestion from the mathlib reviewers based on the \textit{canonical embedding} \lstinline{K →+* ((K →+* ℂ) → ℂ)}. Unlike the naive polynomial approach, this coordinate-free map preserves the indexing of conjugates, effectively embedding $K$ into a Banach algebra $\prod_{\sigma: K \hookrightarrow \mathbb{C}} \mathbb{C}$ equipped with the supremum norm. This architectural choice allowed us to replace ad-hoc polynomial arguments with mathlib’s normed space API, rendering proofs of multiplicativity and the triangle inequality immediate. 

Furthermore, this transition resolved significant administrative friction associated with dependent types. While the naive approach using \texttt{Finset.max'} required an explicit proof term ($h_{\text{max}}$) of non-emptiness as an argument, the canonical embedding is built on the \texttt{Matrix.seminormedAddCommGroup} instance which endows the space of matrices with the properties of a seminormed additive commutative group. In Lean, a matrix is defined as a dependent function (a \texttt{Pi} type) taking row and column indices;  By treating the image of the embedding as a structured matrix (or \texttt{Pi} type), we inherited an algebraic hierarchy. Because this instance provides a total function \lstinline{norm : E → ℝ}, the well-definedness of the house is handled by the type class inference system rather than manual existence proofs. More importantly, this architectural shift allowed us to treat the house not merely as a set-theoretic maximum, but as a norm within a structured algebraic framework, rendering the proofs of subsequent arithmetic bounds immediate.

\begin{lstlisting}[label={lst:params_def},  stepnumber=1, numbersep=5pt, aboveskip=3pt, belowskip=3pt,captionpos=t, caption=Definition of the house]
def house (α : K) : ℝ := ‖canonicalEmbedding K α‖
\end{lstlisting}

We verify that this definition aligns with the mathematical definition $\max_{\sigma} |\sigma(\alpha)|.$ The use of \texttt{Finset.sup'} reflects that the set of embeddings is finite and non-empty. While the definition of \texttt{house} leverages the $L^\infty$-norm of the canonical embedding for its algebraic properties, the following theorem 
explicitly unpacks that norm into the supremum of the absolute values of the 
conjugates.
\begin{lstlisting}[ stepnumber=1, numbersep=5pt, aboveskip=3pt, belowskip=3pt,captionpos=t, caption={Equivalence to the supremum of conjugates}]
theorem house_eq_sup' (α : K) :
    house α = univ.sup' univ_nonempty (fun φ : K →+* ℂ ↦ ‖φ α‖₊) := ...
\end{lstlisting}

\subsection{Siegel's Lemma for Number Fields}
The proof of the Gelfond-Schneider theorem relies on constructing an auxiliary function with coefficients in $\mathcal{O}_K$ that are \enquote{small} in terms of their house. The existence of such coefficients is guaranteed by a version of Siegel's Lemma for algebraic integers.

The classical Siegel's lemma is already part of mathlib and guarantees small integer solutions to underdetermined systems over $\mathbb{Z}.$ 

\begin{theorem}[\texttt{Int.Matrix.exists\_ne\_zero\_int\_vec\_norm\_le'}]
Let $0 < M < N$, and let $a_{jk}$ be rational integers satisfying $|a_{jk}| \leq A$,  
where $A \geq 1$, $1 \leq j \leq M$, and $1 \leq k \leq N.$  Then there exists a set of rational integers $x_1, \ldots, x_N$, not all zero, satisfying $a_{j1}x_1 + \cdots + a_{jN}x_N = 0$ for $1 \leq j \leq M,$ and $|x_k| \leq (NA)^{\frac{M}{N-M}}$ for $1 \leq k \leq N\href{https://github.com/mkaratarakis/mathlib4/blob/9710b82b966813f7664819bc5d25b82d209a05e7/Mathlib/NumberTheory/SiegelsLemma.lean#L214C20-L214C40}{\extlink}.$
\end{theorem}

Our setting requires solutions in $\mathcal{O}_K$ and thus, we formalize Siegel's Lemma for algebraic integers. We achieve this by a standard descent argument: transforming the linear system over $\mathcal{O}_K$ into a larger system over $\mathbb{Z}$ via the following theorem;

\begin{theorem}(\texttt{basis\_repr\_norm\_le\_const\_mul\_house}\href{https://github.com/mkaratarakis/mathlib4/blob/9710b82b966813f7664819bc5d25b82d209a05e7/Mathlib/NumberTheory/NumberField/House.lean#L367}{\extlink})
Let $c$ be a natural number depending on an algebraic number field $K$ of degree $h,$ and its basis $\beta_1, \ldots , \beta_h.$ Then if $\alpha$ is an algebraic integer with $\alpha = a_1\beta_1 + \ldots + a_h\beta_h,$ then $$\left|a _i\right| \leq  c \;\aroof.$$
\end{theorem}

Using the standard integer Siegel Lemma, we obtain a small solution to the flattened system and lift it back to $\mathcal{O}_K.$ The resulting theorem asserts the existence of a \enquote{small} non-zero algebraic integral solution of an
non-trivial underdetermined system of linear equations with algebraic integer coefficients.
\begin{theorem} (\texttt{exists\_ne\_zero\_int\_vec\_house\_le}\href{https://github.com/mkaratarakis/mathlib4/blob/9710b82b966813f7664819bc5d25b82d209a05e7/Mathlib/NumberTheory/NumberField/House.lean#L367}{\extlink})\label{siegel}
Let $0<p<q$, and let $a_{kl}$ be integers with $|a_{kl}|\le A$ ($A\ge1$, $1\le k\le p$, $1\le l\le q$). Then there exist integers $\xi_1,\dots,\xi_q$, not all zero, such that $a_{k1}\xi_1+\cdots+a_{kq}\xi_q=0$ for $1\le k\le p$, and $|\xi_l|<c_1\bigl(1+(c_1 q A)^{\frac{p}{(q-p)}}\bigr)$ for $1\le l\le q.$
\end{theorem}

With the algebraic number theory foundations and Siegel's Lemma fully in place, we now move to the analytic core of the proof of the Gelfond–Schneider Theorem. This phase transitions from algebraic number theory to complex analysis, where the goal is to construct a contradiction based on the growth of an auxiliary function.

\section{Hilbert's seventh problem}
The theorem that we want to prove is the following:
\begin{thm}[Gelfond-Schneider]
Let $\alpha$ and $\beta$ be algebraic numbers with $\alpha \neq 0, 1$ and $\beta$ irrational. Then $\alpha^\beta$ is transcendental.
\end{thm}
The proof has two parts to it and its core rests on the construction of an auxiliary function $R(z)$ that vanishes to a high order at specific points. The proof proceeds by deriving a contradiction between an algebraic lower bound and an analytic upper bound for a value derived from the derivatives of $R(z).$ We now describe in more detail the proof we formalized.

\subsection{Step 1}
Let $\alpha$ and $\beta$ be algebraic numbers with $\alpha\neq 0, 1$ and $\beta$ irrational, and we have to prove that $\alpha^\beta$ is transcendental. The proof proceeds by contradiction. Suppose that $\gamma = \alpha^\beta = e^{\beta \log\alpha}$ (where $log \alpha$ may be any fixed value of the logarithm of $\alpha.$) is also algebraic.

Suppose that $\alpha, \beta, \gamma$ lie in an algebraic field with degree $h=[K:\mathbb{Q}].$ This fact is formalized in a more general version\href{https://github.com/mkaratarakis/mathlib4/blob/9710b82b966813f7664819bc5d25b82d209a05e7/Mathlib/NumberTheory/Transcendental/GelfondSchneider/MainAlg.lean#L62}{\extlink}. In Lean, we bundle these hypotheses into a structure \texttt{Setup}\href{https://github.com/mkaratarakis/mathlib4/blob/9710b82b966813f7664819bc5d25b82d209a05e7/Mathlib/NumberTheory/Transcendental/GelfondSchneider/MainAlg.lean#L105}{\extlink} that also contains the number field $K$ and an embedding $\sigma: K \to \mathbb{C}.$
\begin{lstlisting}[label={lst:setup}, stepnumber=1, numbersep=5pt, aboveskip=3pt, belowskip=3pt,captionpos=t, caption={The structure encoding the Gelfond--Schneider hypotheses.}]
structure Setup where
  (α β : ℂ)
  (K : Type)
  [isField : Field K]
  [isNumberField : NumberField K]
  (σ : K →+* ℂ)
  (α' β' γ' : K)
  hirr : ∀ i j : ℤ, β ≠ i / j
  htriv : α ≠ 0 ∧ α ≠ 1
  hα : IsAlgebraic ℚ α
  hβ : IsAlgebraic ℚ β
  habc : α = σ α' ∧ β = σ β' ∧ α ^ β = σ γ'
  hd : DecidableEq (K →+* ℂ)
\end{lstlisting}


The \texttt{Setup} structure also includes the hypothesis \texttt{habc}, which acts as a dictionary identifying the complex values \lstinline{α,β,α^β} with their algebraic preimages \lstinline{α',β',γ'} in $K.$  This hypothesis bridges the analytic domain ($\mathbb{C}$), where the auxiliary function is defined, and the algebraic domain ($K$), where its coefficients are constructed. Crucially, it ensures that \lstinline{α^β = σ(γ')}, allowing us to apply arithmetic lower bounds to analytic function values.

\subsection*{Step 2: The Linear System and Siegel's Lemma}

Following the reference text, we define parameters $m$ and $n$ dependent on the degree $h$ and a free parameter $q.$ Let $m = 2h + 2,$ and $ n= \frac{q^2}{2m}$ where $q^2 = t$ is a square of a natural number and is a multiple of $2m.$ Also, let $\rho_1, \rho_2, \ldots , \rho_t$ represent the $t$ numbers
$$(a + b\beta)log \alpha, \quad 1 \leq \alpha \leq q, \quad  1 \leq \beta \leq q \href{https://github.com/mkaratarakis/mathlib4/blob/9710b82b966813f7664819bc5d25b82d209a05e7/Mathlib/NumberTheory/Transcendental/GelfondSchneider/MainAlg.lean#L227}{\extlink}.$$
We introduce the auxiliary function
\begin{equation}\label{eq1}
    R(x) = {\eta_1}e^{\rho_1x} + \ldots + {\eta_t}e^{\rho_tx}\href{https://github.com/mkaratarakis/mathlib4/blob/9710b82b966813f7664819bc5d25b82d209a05e7/Mathlib/NumberTheory/Transcendental/GelfondSchneider/MainOrder.lean#L98}{\extlink},
\end{equation}
where the $t = 2mn$ coefficients ${\eta_1} \ldots , {\eta_t} \in \mathcal{O}_K$ are unknowns to be determined by the following conditions. We impose the condition that $R(x)$ vanishes with multiplicity $k$ at the points $x \in \{1, \dots, m\}$ i.e.
\begin{equation}\label{eq2}
(log \alpha)^{-k}  R^{(k)}(l) = 0, \mbox { }0 \leq k \leq n - 1, \mbox{ } 1 \leq l \leq m\end{equation} This yields the system of $mn$ homogeneous linear equations:
$$(\log \alpha)^{-k} R^{(k)}(l) = 0, \quad 0 \le k < n, \quad 1 \le l \le m.$$ The coefficients of this system are numbers in $K$ and 
$$(log \alpha)^{-k}(a + b\beta)log \alpha)^k  e^{l(a+b\beta )log \alpha} = (a + b\beta)^k\alpha^{al}\gamma^{bl},$$
$$1 \leq l \leq m, \mbox{ } 1 \leq a, \mbox{ } b \leq q, \mbox{ } 0\leq k\leq {n-1}.$$

In Lean, we formalize these raw algebraic coefficients as \texttt{systemCoeffs}\href{https://github.com/mkaratarakis/mathlib4/blob/9710b82b966813f7664819bc5d25b82d209a05e7/Mathlib/NumberTheory/Transcendental/GelfondSchneider/MainAlg.lean#L255}{\extlink}. Note that these values lie in the number field $K.$ 

To derive the coefficients ${\eta_1} \ldots , {\eta_t} \in \mathcal{O}_K$ for the auxiliary function~\eqrefcol{eq1} we have to use Siegel's Lemma~\ref{siegel} for algebraic integers. To prepare the linear system for Siegel's Lemma, we must ensure that all coefficients reside in the ring of integers $\mathcal{O}_K$ rather than the field $K.$ This is achieved by clearing denominators using the following lemma:

\begin{lemma}(\texttt{exists\_int\_smul\_isIntegral}\href{https://github.com/mkaratarakis/mathlib4/blob/9710b82b966813f7664819bc5d25b82d209a05e7/Mathlib/NumberTheory/Transcendental/GelfondSchneider/MainAlg.lean#L82}{\extlink})
Let $K$ be a number field and let $\alpha \in K.$ Then there exists a non-zero integer $k \in \mathbb{Z} \setminus \{0\}$ such that $k\alpha$ is an algebraic integer.
\end{lemma} Mathematically, this lemma guarantees the existence of a non-zero integer $c_1 \in \mathbb{Z} \setminus \{0\}$ (formalized as \lstinline{c₁}) such that $c_1 \alpha, c_1 \beta, c_1 \gamma$ are all algebraic integers. By multiplying the entire system by $c_1^{n-1+2mq}$, we clear the denominators introduced by the exponents $k$ (up to $n-1$) and the terms $\alpha^{al}\gamma^{bl}$ (up to exponents of order $mq$) \href{https://github.com/mkaratarakis/mathlib4/blob/9710b82b966813f7664819bc5d25b82d209a05e7/Mathlib/NumberTheory/Transcendental/GelfondSchneider/MainAlg.lean#L336}{\extlink}. In this way, the requirements of the matrix $A$\href{https://github.com/mkaratarakis/mathlib4/blob/9710b82b966813f7664819bc5d25b82d209a05e7/Mathlib/NumberTheory/Transcendental/GelfondSchneider/MainAlg.lean#L346}{\extlink} involved in Siegel's lemma are satisfied as we obtain a matrix $A$ over $\mathcal{O}_K$ with dimensions $mn \times q^2.$ Since we chose $n$ such that $2mn \le q^2,$ the system is undetermined.

We establish a bound on the house of the entries of the matrix $A.$ This result, formalized as \texttt{house\_matrixA\_le}\href{https://github.com/mkaratarakis/mathlib4/blob/cb781672b399c6badb5c70e3f73056bcb31012b0/Mathlib/NumberTheory/Transcendental/GelfondSchneider/MainAlgSetup.lean#L213}{\extlink}, corresponds to the estimate derived in the text where the coefficients satisfy

$${c_2}^n (q + q\overline{|\beta|})^{n-1} \overline{|\alpha|}^{mq} \overline{|\gamma|}^{mq} \leq {c_3}^n n^{\frac{n-1}{2}},$$
with constants $c_2 := |c_1|^{\,2 + 8m^2}$ and 
$c_3 := c_2\,(1 + \roof{\beta'})\, \sqrt{2m}\,
\max\{1,\, \roof{\alpha'}^{2m^2}\, \roof{\gamma'}^{2m^2}
\}$

We then invoke \texttt{exists\_ne\_zero\_int\_vec\_house\_le} (our formalized Siegel's Lemma). This guarantees the existence of a non-trivial set of integer solutions $\eta_1, \ldots, \eta_2$ in $K$ such that
$$\overline{|\eta_k|}\leq c_4^n n^{\frac{(n-1)}{2}}, \quad 1\leq k \leq t,$$
completing the construction of the auxiliary function $R(z).$ We define a new constant $c_4 := \max\{1,\, 2m\, c_1^2\}\, c_3.$ to capture the growth of this solution. Finally, we derive the explicit bound on the coefficients of the auxiliary function. The theorem \lstinline{house_eta_le_c₄_pow}\href{https://github.com/mkaratarakis/mathlib4/blob/cb781672b399c6badb5c70e3f73056bcb31012b0/Mathlib/NumberTheory/Transcendental/GelfondSchneider/MainAlgSetup.lean#L408}{\extlink} asserts that the house of the solution vector is bounded by

$$\overline{|\eta_k|} \leq c_4^n n^{\frac{n+1}{2}}, \quad 1\leq k \leq t.$$

\subsection*{Step 3: Non-vanishing of the auxiliary function $R(x)$}
The next thing we need to prove is the following; since the numbers $\rho_1,\ldots, \rho_t$ are distinct, the fuction $R(x)$ is not identically zero. For suppose otherwise, then on expanding the right hand side of~\eqrefcol{eq1} we have $$\eta_1\rho_1 + \eta_2{\rho_2}^k+\ldots \eta_t {\rho_t}^k=0,$$ a contradiction. Thus, we see from~\eqrefcol{eq2} that 
\begin{equation}\label{eq3}
R(x)= a_{n,l}(x-l)^n + a_{n+1,l}(x-l)^{n+1}+\cdots, \quad 1\leq l \leq m,
\end{equation}
where $a_{n,l},a_{n+1,l},\ldots$ are not all zero. Hence, there must be a natural number $r$ such that $R^{(k)}(l)=0,$ $0\leq k \leq {r-1},$ $1\leq l \leq m.$ But for $1\leq l_0\leq m$ we have $R^{(r)}(l_0)\neq 0$ so that we see from~\eqrefcol{eq3} that $r\geq n.$

Before defining the order of vanishing, we must establish that the auxiliary function $R(z)$ is not identically zero. In the classical proof, this relies on the distinctness of the exponents $\rho_t.$ We first prove that the map $t \mapsto \rho_t$ is injective\href{https://github.com/mkaratarakis/mathlib4/blob/cb781672b399c6badb5c70e3f73056bcb31012b0/Mathlib/NumberTheory/Transcendental/GelfondSchneider/MainOrder.lean#L47}{\extlink} for distinct indices. This allows us to construct a Vandermonde matrix \texttt{V}\href{https://github.com/mkaratarakis/mathlib4/blob/cb781672b399c6badb5c70e3f73056bcb31012b0/Mathlib/NumberTheory/Transcendental/GelfondSchneider/MainOrder.lean#L87}{\extlink} based on the values $\rho_1\ldots \rho_t.$ We then prove that the determinant of $V$ is non-zero, which establishes the linear independence of the exponentials. Since Siegel's Lemma guarantees that the coefficients $\eta_1 \ldots \eta_t$ are non-zero, the linear independence implies that the linear combination $R(z)$ cannot vanish everywhere\href{https://github.com/mkaratarakis/mathlib4/blob/cb781672b399c6badb5c70e3f73056bcb31012b0/Mathlib/NumberTheory/Transcendental/GelfondSchneider/MainOrder.lean#L164}{\extlink}.

Since $R(z)$ is a non-zero analytic function, it must have a finite order of vanishing at any point. We identify and define the minimal order $r$ across the set of points $\{1, \dots, m\}.$ In pen-and-paper proofs, one typically asserts the existence of a \enquote{first non-vanishing derivative} $r.$ In Lean, we formalize this constructively using \texttt{analyticOrderAt}, which returns the order of the zero of an analytic function at a point.

\begin{lstlisting}[caption={Selecting the minimal order}]
lemma exists_min_analyticOrderAt :
  let s : Finset (Fin (h7.m)) := Finset.univ
  ∃ l₀' ∈ s, (∃ y, (analyticOrderAt (h7.R q hq0 h2mq) (l₀' + 1)) = y ∧
  (∀ (l' : Fin (h7.m)), l' ∈ s → 
  y ≤ (analyticOrderAt (h7.R q hq0 h2mq) (l' + 1)))) := ...
\end{lstlisting}
We define \lstinline{r}\href{https://github.com/mkaratarakis/mathlib4/blob/cb781672b399c6badb5c70e3f73056bcb31012b0/Mathlib/NumberTheory/Transcendental/GelfondSchneider/MainAnalytic.lean#L91}{\extlink} as this minimum value and \lstinline{l₀'}\href{https://github.com/mkaratarakis/mathlib4/blob/cb781672b399c6badb5c70e3f73056bcb31012b0/Mathlib/NumberTheory/Transcendental/GelfondSchneider/MainOrder.lean#L259}{\extlink}as the point where this minimum is attained using \texttt{Classical.choose}. We then prove $r \ge n$\href{https://github.com/mkaratarakis/mathlib4/blob/cb781672b399c6badb5c70e3f73056bcb31012b0/Mathlib/NumberTheory/Transcendental/GelfondSchneider/MainPostAnalytic.lean#L132}{\extlink}, mirroring the deduction in the text that the system of linear equations forces the first $n$ derivatives to vanish at all points. 

\section*{Step 5: The Arithmetic Lower Bound}
We examine the normalized derivative value
\begin{equation}\label{eq:4}
\rho = (\log \alpha)^{-r} R^{(r)}(l_0)\neq 0    
\end{equation}
The key idea is that $\rho$ is simultaneously algebraic (so its house cannot be too small) and analytically small (so its absolute value can be made very small by taking $r$ large). 

Expanding the $r$-th derivative of $R(z)$, we see that $R^{(r)}(l_0)$ is a linear combination of terms of the form
$$\eta_t (a+b\beta)^r \alpha^{a l_0} \gamma^{b l_0} (\log \alpha)^r.$$
In Lean, we define these expressions as \texttt{systemCoeffs\_r}\href{https://github.com/mkaratarakis/mathlib4/blob/cb781672b399c6badb5c70e3f73056bcb31012b0/Mathlib/NumberTheory/Transcendental/GelfondSchneider/MainAnalytic.lean#L109}{\extlink} and their total sum is $$\texttt{rho} := \sum_{t=1}^{q^2}, \eta_t (a+b\beta)^r \alpha^{a l_0} \gamma^{b l_0} (\log \alpha)^r \href{https://github.com/mkaratarakis/mathlib4/blob/cb781672b399c6badb5c70e3f73056bcb31012b0/Mathlib/NumberTheory/Transcendental/GelfondSchneider/MainAnalytic.lean#L130}{\extlink}.$$

In our formalization, we distinguish between the analytic definition involving the derivative (formalized as \lstinline{ρᵣ} $:=(\log \alpha)^{-r}
R^{(r)}(l_0) \href{https://github.com/mkaratarakis/mathlib4/blob/cb781672b399c6badb5c70e3f73056bcb31012b0/Mathlib/NumberTheory/Transcendental/GelfondSchneider/MainAnalytic.lean#L130C5-L130C7}{\extlink}$ 
and its algebraic expansion (formalized as \texttt{rho} above). We then prove that these two definitions coincide.

A subtle technical detail in establishing this equality is the alignment of the evaluation point $l_0.$ In Lean, we identify the index of the minimal vanishing order as a term $l_0' : \texttt{Fin } m.$ Since \texttt{Fin } m is canonically $0$-indexed (representing $i < m$), the actual complex point where the derivative is evaluated is $l_0' + 1.$ This shift is crucial for the algebraic expansion: the exponent $l_0$ appearing in the term $\alpha^{a l_0} \gamma^{b l_0}$ corresponds strictly to this shifted value, $l_0' + 1$, ensuring that the analytic evaluation at the lattice point matches the algebraic coefficients derived from the system.


\section*{Step 6 : Bounding $\rho$ algebraically} 
The number $\rho$ lies in $K.$ Thus, similarly to the previous section, we multiply $\rho$ by ${c_1}^{r+2mq}\rho$ to obtain an algebraic integer \texttt{$c_\rho$} in $K.$ Since $r$ was chosen as the \textit{minimal} order of vanishing, we know that $\rho \neq 0.$ This non-vanishing, combined with integrality, allows us to apply the norm inequality $|N_{K/\mathbb{Q}}(c_\rho \rho)| \ge 1$, which provides the arithmetic lower bounds required for the contradiction:

\begin{equation}\label{eq5}
|N(\rho)| > c_1^{-h(r+2mq)} > c_5^{-r}\href{https://github.com/mkaratarakis/mathlib4/blob/7e18c3edf49afd671b6af4535839428570e454ce/Mathlib/NumberTheory/Transcendental/GelfondSchneider/MainPostAnalytic.lean#L156C29-L156C30}{\extlink}
\end{equation}
where norm $N(\rho)$ is the product of the principal value $|\rho|$ and its conjugates.

On the other hand
\begin{equation}\label{eq6}
\roof{\rho}\leq t{c_4}^nn^{{(n-1)}{2}}{(c_6q)^r}{c_7}^q\leq {c_8}^r{r^{r+\frac{3}{2}}}\href{https://github.com/mkaratarakis/mathlib4/blob/7e18c3edf49afd671b6af4535839428570e454ce/Mathlib/NumberTheory/Transcendental/GelfondSchneider/MainAnalyticBounds.lean#L702}{\extlink}.
\end{equation}
with the following constants:
$$\begin{aligned}
c_5 &:= \bigl(|c_1| + 1\bigr)^{\,h\, (1 + 4\, m^2)}, \\[2mm]
c_6 &:= |c_1| \,\bigl(1 + \roof{\beta'}\bigr), \\[1mm]
c_7 &:= \Bigl( |c_1| \cdot |c_1| \cdot \bigl(|c_1| \cdot (\roof{\alpha'} \cdot (|c_1| \cdot \roof{\gamma'}))\bigr) \Bigr)^{m}, \\[1mm]
c_8 &:= c_6 \,\sqrt{2\, m}\, c_7^{2\, m} \, c_4 \, 2\, m.
\end{aligned}$$
To derive a contradiction, we must also establish an analytic upper bound for $\rho.$

\section*{Step 7: The Analytic Upper Bound}
The proof continues with bounding $\rho$ analytically. This is achieved via a helper function $S(z)$ whose construction constitutes the analytic heart of the Gelfond–Schneider proof. This global auxiliary function is defined by a product involving $R$ and poles at finitely many lattice points and is given in the text as:

$$S(z) = r! \frac{R(z)}{(z-\ell_0)^r}
\prod_{\substack{k = 1 \\ k \neq \ell_0}}^m
\left( \frac{\ell_0 - k}{z - k} \right)^{r}.$$

The central engineering issue in our formalization is how to represent the auxiliary function \(S(z)\). In the classical proof, \(S\) is introduced by a formula that is implicitly \emph{meromorphic}: it involves the factor
$$\frac{R(z)}{(z-\ell_0)^r}
\prod_{\substack{k = 1 \\ k \neq \ell_0}}^m
\left(\frac{1}{z - k} \right)^{r},$$
which has poles at each integer \(1,\dots,m\). On paper, these poles are harmless because one simply avoids them when applying Cauchy's integral formula. In Lean, however, a meromorphic definition would propagate partiality everywhere: one would need to encode domains of definition, track that the Cauchy contour avoids all poles, and prove the absence of singularities in every analytic estimate. This would significantly enlarge the proof state, introduce layers of option types or subtypes, and complicate every downstream lemma involving \(S\). For this reason, we deliberately do \emph{not} formalize \(S\) as a meromorphic expression. Instead, we construct \(S\) as a \emph{total, piecewise holomorphic} function using a case split on whether the input lies on one of the integers \(1,\dots,m\). 

\paragraph*{The Structure of $R $ and Its Local Factors}
Formalizing $S(z)$ requires handling removable singularities. First, we formalize the removability of singularities in the auxiliary function $R.$ We construct a globally defined, total function $R'(z)$ that represents the quotient $R(z)(z-l)^{-r}$ everywhere on $\mathbb{C}.$
While mathematically this is often treated implicitly, in Lean we must explicitly construct the function values at the singular points $l+1.$ We achieve this through a patching strategy involving three components. First, using the fact that $R(z)$ vanishes to order $r$ at $l,$ we invoke the following lemma to extract a local holomorphic factor $R'_U(z)$\href{https://github.com/mkaratarakis/mathlib4/blob/9710b82b966813f7664819bc5d25b82d209a05e7/Mathlib/NumberTheory/Transcendental/GelfondSchneider/MainHol.lean#L104}{\extlink} valid in some neighborhood $U$\href{https://github.com/mkaratarakis/mathlib4/blob/9710b82b966813f7664819bc5d25b82d209a05e7/Mathlib/NumberTheory/Transcendental/GelfondSchneider/MainHol.lean#L106}{\extlink} of $l.$

\begin{theorem}[\texttt{exists\_analyticOn\_factor\_R\_at\_add\_one}]
There exists a neighborhood $U \subset \mathbb{C}$ of $l$ and a function $R'_U : \mathbb{C} \to \mathbb{C}$ analytic on $U$ such that $\forall z \in U, \mbox{ } R(z) = (z - l)^r \, R'_U(z),$ with $R'(l) \neq 0.$
\end{theorem}

Second, we define the standard rational form $R|_{\mathbb{C} \setminus \{l\}}(z)=R(z)(z-l)^{-r}$\href{https://github.com/mkaratarakis/mathlib4/blob/9710b82b966813f7664819bc5d25b82d209a05e7/Mathlib/NumberTheory/Transcendental/GelfondSchneider/MainHol.lean#L116}{\extlink} on the punctured plane $\mathbb{C}\setminus\{l\}.$ Finally, we define the global function $R'(z)$ by gluing these definitions:

$$R'(z) = 
\begin{cases} 
R'_U(z) & \text{if } z = l, \\
R|_{\mathbb{C} \setminus \{l\}}(z) & \text{if } z \neq l.
\end{cases}\href{https://github.com/mkaratarakis/mathlib4/blob/9710b82b966813f7664819bc5d25b82d209a05e7/Mathlib/NumberTheory/Transcendental/GelfondSchneider/MainHol.lean#L119}{\extlink}
$$

We then prove that $R'$ is analytic everywhere by showing it agrees with the analytic function $R'_U$ on the neighborhood U and with the analytic product $R|_{\mathbb{C} \setminus \{\lambda\}}$ on the complement of $U$, thereby formalizing the analytic continuation required for the proof \href{https://github.com/mkaratarakis/mathlib4/blob/e805edf67278fdd61da7bda2bad20d8f6f0cf6f8/Mathlib/NumberTheory/Transcendental/GelfondSchneider/MainHol.lean#L124}{\extlink}.

So now, to formalize the auxiliary function $S(z)$ as a total, everywhere-analytic function, we extend the patching strategy used for $R(z)$ to handle multiple removable singularities. Let $\mathcal{P}=\{1,\ldots,m\}\href{https://github.com/mkaratarakis/mathlib4/blob/e805edf67278fdd61da7bda2bad20d8f6f0cf6f8/Mathlib/NumberTheory/Transcendental/GelfondSchneider/MainHol.lean#L166}{\extlink}$ be the set of singularities.

The global function $S(z)$ is then constructed by gluing together three distinct analytic components. On the complement of the singularities $(\mathbb{C}\setminus \mathcal{P})$, we use the standard formula from the text:

$$S|_{\mathbb{C}\setminus \mathcal{P}}(z):= r! \frac{R(z)}{(z-\ell_0)^r}
\prod_{\substack{k = 1 \\ k \neq \ell_0}}^m
\left( \frac{\ell_0 - k}{z - k} \right)^{r}\href{https://github.com/mkaratarakis/mathlib4/blob/e805edf67278fdd61da7bda2bad20d8f6f0cf6f8/Mathlib/NumberTheory/Transcendental/GelfondSchneider/MainHol.lean#L191}{\extlink}.$$

At the critical point $l_0$, we define a specific local function $S_{l_0}(z)$ which uses the previously derived local factor $R'(z)$ to eliminate the pole of order $r:$

$$S_{l_0}(z) := r! R'(z) \prod_{\substack{k = 1 \\ k \neq \ell_0}}^m\left( \frac{l_0 - k}{z - k} \right)^r\href{https://github.com/mkaratarakis/mathlib4/blob/e805edf67278fdd61da7bda2bad20d8f6f0cf6f8/Mathlib/NumberTheory/Transcendental/GelfondSchneider/MainHol.lean#L223}{\extlink}.
$$

Similarly, for any other point $k\in (\mathcal{P}\setminus \{l_0\}),$ we define a local function $$
S_l(z) =  r! R'(z)(z - l_0)^{-r}  (l_0 - l)^r\prod_{\substack{k \notin \{l_0, l\}}}^m \left( \frac{l_0 - k}{z - k} \right)^r\href{https://github.com/mkaratarakis/mathlib4/blob/e805edf67278fdd61da7bda2bad20d8f6f0cf6f8/Mathlib/NumberTheory/Transcendental/GelfondSchneider/MainHol.lean#L228}{\extlink}
$$that regularizes the singularity at $k\in (\mathcal{P}\setminus \{l_0\}),$ using its local factor $R'(z),$ while preserving the singular term associated with $l_0.$ 

The global auxiliary function $S(z)$\href{https://github.com/mkaratarakis/mathlib4/blob/e805edf67278fdd61da7bda2bad20d8f6f0cf6f8/Mathlib/NumberTheory/Transcendental/GelfondSchneider/MainHol.lean#L236}{\extlink} is then defined piecewise as follows:

$$S(z) = 
\begin{cases} 
S_{l_0}(z) & \text{if } z = l_0, \\
S_k(z) & \text{if } z = k \text{ for } k \in \mathcal{P} \setminus \{l_0\}, \\
S|_{\mathbb{C}\setminus \mathcal{P}}(z) & \text{if } z \notin \mathcal{P}.
\end{cases}
$$

Finally, we need to establish the global analyticity of $S(z).$ Each of these three components is a separate analytic object with its own proof of analyticity and proving the global analyticity of $S(z)$ requires establishing that these piecewise definitions stitch together into a single holomorphic function. For points $z$ in the complement of the singularities $\mathbb{C}\setminus{\mathcal{P}}$, analyticity follows directly from the composition of standard analytic functions in the product formula $S|_{\mathbb{C}\setminus{\mathcal{P}}}(z)\href{https://github.com/mkaratarakis/mathlib4/blob/e805edf67278fdd61da7bda2bad20d8f6f0cf6f8/Mathlib/NumberTheory/Transcendental/GelfondSchneider/MainHol.lean#L216}{\extlink}.$ 

At the removable singularities $l\in \mathcal{P},$ the proof relies on the local analytic behavior of the repaired components $S_{l_0}(z)$ and $S_k(z).$ We first establish that these local functions are analytic in a ball of radius 1 around their respective centers (\href{https://github.com/mkaratarakis/mathlib4/blob/e805edf67278fdd61da7bda2bad20d8f6f0cf6f8/Mathlib/NumberTheory/Transcendental/GelfondSchneider/MainHol.lean#L477}{\extlink} and \href{https://github.com/mkaratarakis/mathlib4/blob/e805edf67278fdd61da7bda2bad20d8f6f0cf6f8/Mathlib/NumberTheory/Transcendental/GelfondSchneider/MainHol.lean#L420}{\extlink}), a fact derived from the everywhere-analyticity of the underlying factor $R'(z).$ 

The final glueing step, formalized in \href{https://github.com/mkaratarakis/mathlib4/blob/e805edf67278fdd61da7bda2bad20d8f6f0cf6f8/Mathlib/NumberTheory/Transcendental/GelfondSchneider/MainHol.lean#L507}{\extlink}, verifies that these local definitions coincide with the global naive form on the punctured neighborhoods of the singularities (lemmas \href{https://github.com/mkaratarakis/mathlib4/blob/e805edf67278fdd61da7bda2bad20d8f6f0cf6f8/Mathlib/NumberTheory/Transcendental/GelfondSchneider/MainHol.lean#L246}{\extlink} and \href{https://github.com/mkaratarakis/mathlib4/blob/e805edf67278fdd61da7bda2bad20d8f6f0cf6f8/Mathlib/NumberTheory/Transcendental/GelfondSchneider/MainHol.lean#L290}{\extlink}). By demonstrating this local agreement and the regularity of each patch, we rigorously prove that $S(z)$ is analytic at every point in the complex plane.

\section*{Step 7 : Employing $S(z)$}
We now determine a suitable upper bound for $\mid \rho \mid.$ We will apply Cauchy's integral formula to the global function 
$$S(z) = r! \frac{R(z)}{(z-\ell_0)^r}
\prod_{\substack{k = 1 \\ k \neq \ell_0}}^m
\left(
\frac{\ell_0 - k}{z - k}
\right)^{r}.$$

Then we have \begin{equation}\label{eq7}
\rho= (log \alpha)^{-r}S(l_0)=(\log \alpha)^{-r} \frac{1}{2\pi i}
\oint_{C} \frac{S(z)}{z - \ell_0}dz
\end{equation} where $C$ is the circle $|z| = m(1+\frac{r}{q}),$ so that $l_0$ $(\leq m)$ lies inside $C\href{https://github.com/mkaratarakis/mathlib4/blob/e805edf67278fdd61da7bda2bad20d8f6f0cf6f8/Mathlib/NumberTheory/Transcendental/GelfondSchneider/MainHol.lean#L622}{\extlink}.$

To evaluate the integral, we observed that the contour $C$ strictly encloses the set of singularities $\mathcal{P}.$ Consequently, for all $z$ on the contour, the global function $S(z)$ reduces simply to $S(z) = S|_{\mathbb{C}\setminus \mathcal{P}}(z)\href{https://github.com/mkaratarakis/mathlib4/blob/e805edf67278fdd61da7bda2bad20d8f6f0cf6f8/Mathlib/NumberTheory/Transcendental/GelfondSchneider/MainHol.lean#L650}{\extlink}.$ This simplification permits us to apply the analytic upper bounds derived for the product form directly to the integrand.



We now obtain a bound for $|R(z)|.$ Using the bounds on the coefficients $\eta_k$ derived from Siegel's Lemma, we prove that as $z$ varies on the circle we have 

$$|R(z)| 
\le t \max_{1 \le k \le t} |\eta_k|\,
e^{(q + q|b|)\left(\log|\alpha|\, m \left(1 + \frac{r}{q}\right)\right)}\\
\le t\, c_4^n n^{\frac{n+1}{2}} c_9^{r+q} \le c_{10}^r r^{\frac{r+3}{2}}\href{https://github.com/mkaratarakis/mathlib4/blob/e805edf67278fdd61da7bda2bad20d8f6f0cf6f8/Mathlib/NumberTheory/Transcendental/GelfondSchneider/MainBounds.lean#L58}{\extlink}.$$
The constants that we had to define were the following: $$
\begin{aligned}
c_9 &:= \exp\left(|1 + |\beta|| \cdot |\log \alpha| \cdot m\right) \\
c_{10} &:= 2m c_4 c_9^{1 + 2m}
\end{aligned}
$$

The function $S(z)$ contains a product term $\prod |(\ell_0 - k)/(z - k)|^r.$ Using the geometric lower bounds established in step 1, this lemma provides an upper bound for this product term.
This specific radius is chosen to strictly enclose the set of singularities $\mathcal{P}={1\ldots,m}$ while maintaining a quantifiable distance from them (formalized in \href{https://github.com/mkaratarakis/mathlib4/blob/e805edf67278fdd61da7bda2bad20d8f6f0cf6f8/Mathlib/NumberTheory/Transcendental/GelfondSchneider/MainBounds.lean#L49}{\extlink}). 

We establish a uniform geometric separation result 
$$
\left| z - \ell_0 \right| \ge \left| z \right| - \left| \ell_0 \right| \ge m \left(1 + \frac{r}{q}\right) - m = \frac{mr}{q}\href{https://github.com/mkaratarakis/mathlib4/blob/e805edf67278fdd61da7bda2bad20d8f6f0cf6f8/Mathlib/NumberTheory/Transcendental/GelfondSchneider/MainBounds.lean#L297}{\extlink},$$ and


$$
|z - k| \ge \frac{mr}{q}, \quad 1 \le k \le m\href{https://github.com/mkaratarakis/mathlib4/blob/e805edf67278fdd61da7bda2bad20d8f6f0cf6f8/Mathlib/NumberTheory/Transcendental/GelfondSchneider/MainBounds.lean#L320}{\extlink},
$$
proving that for any $z\in C$ and any integer $k\in \mathcal{P},$ the distance is bounded from below by the difference between the radius and the largest singularity $m.$

This lower bound on the distance ensures that the interpolation nodes do not cause the function to diverge on the contour. Applying this estimate to the rational component of the auxiliary function, we obtain the following upper bound for the product term:

$$\left| (z - \ell_0)^{-r} \prod_{\substack{k=1 \\ k \neq \ell_0}}^{m} 
\left( \frac{\ell_0 - k}{z - k} \right)^r \right|
\;\le\; c_{11}^r \left( \frac{q}{r} \right)^{m r}\href{https://github.com/mkaratarakis/mathlib4/blob/e805edf67278fdd61da7bda2bad20d8f6f0cf6f8/Mathlib/NumberTheory/Transcendental/GelfondSchneider/MainBounds.lean#L341}{\extlink}.$$

Combining the geometric estimates, we establish the bound for $|S(z)|$ on the integration contour $C.$ We define a composite constant $c_{12} = (2m)^{m/2} c_{10} c_{11}$ and prove that 

$$|S(x)| \leq r!{c^r_{10}}{\frac{r(r+3)}{2}}{c^r_{11}}(\frac{q}{r})^{mr}\leq {c^r_{12}}r^{r\frac{(3-r)+3}{2}}\href{https://github.com/mkaratarakis/mathlib4/blob/e805edf67278fdd61da7bda2bad20d8f6f0cf6f8/Mathlib/NumberTheory/Transcendental/GelfondSchneider/MainBounds.lean#L538}{\extlink}$$

which asserts that $|S(z)|$ decays as $r$ increases.
Applying the integral formula on the contour $C$ \eqref{eq7} and substituting the bound for $|S(z)|$, we derive the definitive analytic upper bound for $\rho:$
\begin{equation}\label{eq8}
\begin{aligned}
|\rho| &\;\le\; \frac{1}{2\pi} \left| (\log \alpha)^{-r} \right| 
\left|\oint_{C} \frac{S(z)}{z - \ell_0} \right|  \left|dz \right| \\[2mm]
&\;\le\; |(\log \alpha)^{-r}| \, m \left(1 + \frac{r}{q}\right) c^r_{12} r^{\frac{r(3-m) + 3}{2}} \frac{q}{mr}
\;\le\; c_{13}^r r^{\frac{r(3-r)+3}{2}}\href{https://github.com/mkaratarakis/mathlib4/blob/e805edf67278fdd61da7bda2bad20d8f6f0cf6f8/Mathlib/NumberTheory/Transcendental/GelfondSchneider/MainBounds.lean#L813}{\extlink}
\end{aligned}
\end{equation} which is an analytic estimate independent of the algebraic norm.

\subsection*{Step 8: Lower bound and contradiction}
We now switch perspectives again from complex analysis to algebraic number theory. To establish the contradiction, we compute an upper bound for the norm of $\rho$ over $\mathbb{Q}.$ From~\eqrefcol{eq6} and~\eqrefcol{eq8} we have
$$|N(\rho)| \le c^r_{14} \, r^{(h-1)\left(r + \frac{3}{2} + \frac{(3-m)r + 3}{2}\right)}\href{https://github.com/mkaratarakis/mathlib4/blob/e805edf67278fdd61da7bda2bad20d8f6f0cf6f8/Mathlib/NumberTheory/Transcendental/GelfondSchneider/MainBounds.lean#L988}{\extlink},$$ with constants:

$$\begin{aligned}
c_{13} &= \left( |\log \alpha|^{-1} + 1 \right) m \left( 2 + \frac{1}{m} \right) c_{12}, \\
c_{14} &= c_8^{h-1} c_{13}.
\end{aligned}$$
Replacing $m$ by $2h + 2$ we now have
$$|N(\rho)| \le c^r_{14} \, r^{\frac{3h-r}{2}}
$$
and from~\eqrefcol{eq5} we deduce that
$$r^{\frac{3h-r}{2}}<c^r_{14}c^r_{5}=c^r_{15}\href{https://github.com/mkaratarakis/mathlib4/blob/e805edf67278fdd61da7bda2bad20d8f6f0cf6f8/Mathlib/NumberTheory/Transcendental/GelfondSchneider/MainBounds.lean#L1108}{\extlink}$$


For sufficiently large $r$ (specifically $r \ge n$), the term $r^{r/2}$ on the left grows super-exponentially, while $c_{15}^r$ grows only exponentially. Thus, the inequality cannot hold for large $n.$ In the formalization, we explicitly choose $q$ sufficiently large (defined as $q = 12mh \, \lceil c_{15}^{4} \rceil$) to force this contradiction.


This culminates in the formal proof of the Gelfond-Schneider theorem. The theorem requires $\alpha, \beta$ to be algebraic, with $\alpha \notin \{0, 1\}$ and $\beta$ irrational \href{https://github.com/mkaratarakis/mathlib4/blob/e805edf67278fdd61da7bda2bad20d8f6f0cf6f8/Mathlib/NumberTheory/Transcendental/GelfondSchneider/statement.lean#L24}{\extlink}.

\begin{lstlisting}[caption={The Gelfond-Schneider Theorem}]
/-- The Gelfond-Schneider Theorem (Hilbert's Seventh Problem). -/
theorem transcendental_cpow_of_isAlgebraic_of_irrational (α β : ℂ)
    (hα : IsAlgebraic ℚ α) (hβ : IsAlgebraic ℚ β)
    (htriv : α ≠ 0 ∧ α ≠ 1) (hirr : ∀ i j : ℤ, β ≠ i / j) :
    Transcendental ℚ (α ^ β)
\end{lstlisting}


As a sanity check for our formalization of the Gelfond-Schneider theorem, we proved the transcendence of the Gelfond-Schneider constant, $\sqrt{2}^{\sqrt{2}}$\href{https://github.com/mkaratarakis/mathlib4/blob/e805edf67278fdd61da7bda2bad20d8f6f0cf6f8/Mathlib/NumberTheory/Transcendental/GelfondSchneider/statement.lean#L199}{\extlink}.

\begin{lemma}
The number $\sqrt{2}^{\sqrt{2}}$ is transcendental over $\mathbb{Q}.$
\end{lemma}


\section{Discussion}
The formalization process necessitated a rigorous treatment of several concepts often glossed over in classical pen-and-paper arguments, particularly those found in standard texts such as Hua~\citeref{hua1982}.

Classical proofs in Diophantine approximation often rely on $O(\cdot)$ notation or generic constants. To satisfy the strict requirements of the proof assistant, we explicitly constructed a sequence of constants $c_1, \dots, c_{15}$ dependent on the degree of the number field and the geometric properties of the embedding. This explicit tracking was crucial for the final contradiction, where we demonstrated that the analytic upper bound grows slower than the algebraic Liouville lower bound. This aligns with the effective methods later developed by Baker~\citeref{baker1966}, though our focus here was qualitative transcendence.

A significant challenge in the analytic portion of the proof was the construction of the global auxiliary function $S(z).$ Classically treated as a quotient with removable singularities at integer points (Chapter 17, \citeref{hua1982}), in dependent type theory this required the construction of a piecewise total function. We utilized local power series expansions to define the function values at the singularities $l \in \{1, \dots, m\}$ and glued these to the rational function definition valid on the complement. Proving the global analyticity of this stitched function was a non-trivial exercise in formal complex analysis, leveraging Mathlib's analytic function library~\citeref{mathlib2020}.


\section{Conclusion}

We have presented a complete formalization of the Gelfond--Schneider theorem in Lean, verifying that for algebraic numbers $\alpha \neq 0, 1$ and irrational algebraic $\beta$, the value $\alpha^\beta$ is transcendental. This result, originally established independently by Gelfond~\citeref{gelfond1934} and Schneider~\citeref{schneider1935}, solved Hilbert's Seventh Problem. As a concrete application of our formal framework, we formally derived the transcendence of the Gelfond--Schneider constant $\sqrt{2}^{\sqrt{2}}.$

This work lays the foundation for more advanced formalizations in transcendental number theory. The most natural extension is Baker's Theorem on linear forms in logarithms~\citeref{tnt}, which generalizes Gelfond--Schneider to arbitrary finite sets of logarithms. While the overarching strategy -- auxiliary functions and extrapolation -- remains similar, Baker's theorem introduces significant complexity regarding multi-variable analysis and sharper constant management.

The algebraic structure of our proof is largely independent of the complex embedding. By replacing the complex analytic lemmas (specifically Cauchy's integral formula) with their $p$-adic counterparts (Schnirelmann integrals), one could formalize the $p$-adic Gelfond--Schneider theorem established by Mahler~\citeref{mahler1935} and Veldkamp~\citeref{veldkamp1941}.

Additionally, the explicit nature of our constant handling suggests that quantitative transcendence measures (effective lower bounds on $|\alpha^\beta - p/q|$) are within reach. Such bounds, refined by Feldman~\citeref{feldman1968}, require the precise type of constant tracking we have implemented in this work.

Our formalization currently focuses on the classical exponential function, which acts as the bridge between the additive and multiplicative structures of complex numbers. A natural generalization is to extend this framework to elliptic curves. In this setting, the \emph{Weierstrass $\wp$-function} serves as the analogue to the exponential function, mapping the linear complex plane onto the algebraic curve. Formalizing this result would require interfacing our analytic estimates with the algebraic group operations of elliptic curves and the properties of the Weierstrass $\wp$-function~\citeref{elliptic} formalized in Lean~\citeref{angdinata}.

Furthermore, the analytic techniques formalized here -- specifically the construction of auxiliary functions and the application of Siegel's Lemma -- are central to the proof of the Schneider--Lang theorem~\citeref{lang1966}. This theorem generalizes the Gelfond--Schneider result from the exponential function to meromorphic functions satisfying algebraic differential equations, providing a pathway to formalizing transcendence results for abelian varieties. 

From a verification perspective, the explicit constants derived in our proof suggest the possibility of implementing decision procedures~\citeref{decision} for specific classes of transcendental numbers. By combining these bounds with interval arithmetic, one could automate the verification of linear independence for specific sets of logarithms, automating a tedious class of proofs in number theory.

Finally, this work contributes to the library of facts supporting the conditional verification of Schanuel's Conjecture~\citeref{conjecture}. While Schanuel's Conjecture remains unproven, formalizing its consequences -- which include the Gelfond--Schneider theorem as a special case -- provides a rigorous framework for exploring the algebraic independence of values of the exponential function.

\bibliography{references}

\end{document}